\documentclass[english,twocolumn,aps,crop,tikz,superscriptaddress]{revtex4-1}
\usepackage[utf8]{inputenc}
\usepackage[T1]{fontenc}
\usepackage{babel}
\usepackage{amsmath}
\usepackage{graphicx}
\usepackage{fancyhdr}
\usepackage{braket}
\usepackage{pagenote}
\usepackage{siunitx}
\usepackage{xcolor}
\usepackage{lipsum}

\usepackage[colorlinks=true,urlcolor=blue,citecolor=blue,linkcolor=blue]{hyperref}

\makepagenote

\begin{document}

\title{Simulating the dynamics of braiding of Majorana zero modes using an IBM quantum computer}

\author{John P. T. Stenger}
\affiliation{Department of Physics and Astronomy, University of Pittsburgh, Pittsburgh, PA 15260, USA}

\author{Nicholas T. Bronn}
\affiliation{IBM Quantum, IBM T.J. Watson Research Center, Yorktown Heights, New York 10598, USA}

\author{Daniel J. Egger}
\affiliation{IBM Quantum, IBM Research – Zurich, Säumerstrasse 4, 8803 Rüschlikon, Switzerland}

\author{David Pekker}
\affiliation{Department of Physics and Astronomy, University of Pittsburgh, Pittsburgh, PA 15260, USA}

\date{\today}

\begin{abstract}
We simulate the dynamics of braiding Majorana zero modes on an IBM Quantum computer. We find the native quantum gates introduce too much noise to observe braiding. Instead, we use Qiskit Pulse to develop scaled two-qubit quantum gates that better match the unitary time evolution operator and enable us to observe braiding. This work demonstrates that quantum computers can be used for simulation, and highlights the use of pulse-level control for programming quantum computers and constitutes the first experimental evidence of braiding via dynamical Hamiltonian evolution. 
\end{abstract}

\maketitle

\section{Introduction}
Quantum computers may significantly outperform classical ones in the area of simulation of quantum systems~\cite{Feynman1982,Lloyd1996} and other specialized algorithms~\cite{Boixo2018, Shor1997}.  However, we are currently in the era of noisy quantum computing~\cite{Preskill2018} where only a small number of qubits, with relatively short coherences times, can be entangled. Fortunately, recent results demonstrating hardware specific optimization~\cite{Shi2019,Glaser2015,Leung2017} and quantum advantage with short-depth noisy circuits~\cite{Bravyi2020} offer hope for the usefulness of near-term systems.
Here, we simulate a quantum topological condensed matter system on an IBM Quantum processor within the qubits' coherence times using pulse-level instructions provided by Qiskit Pulse~\cite{Qiskit2019,McKay2018, Alexander2020}. Ideally, the simulation would occur by a continuous time-evolution of the qubits under the appropriate spin Hamiltonian obtained from a transformation of the fermion Hamiltonian. Practically, this ``analog'' simulation must be decomposed and mapped onto the calibrated native basis gates of a quantum computer, making it ``digital''. This digital implementation on noisy quantum hardware limits the flexibility needed to avoid the accumulation of unnecessary errors. Here, we demonstrate a ``semi-analog'' approach to noise reduction: a pulse-scaling technique that, without additional calibration~\cite{Alexander2020, Garion2020}, gets us closer to the ideal analog simulation.

Topologically protected quantum computation works by moving nonabelian anyons, such as Majorana zero modes (MZMs), around each other in two dimensions to form three dimensional braids in space-time~\cite{Kitaev2001,Nayak2008,Zeng2018}. This approach is advantageous as it offers protection from local perturbations.
The quest for topological quantum computation has focused on hybrid superconducting-semiconducting~\cite{Alicea2012,Beenakker2013,Chen2017,Deng2016,Kitaev2001,Lutchyn2010, Mourik2012,Oreg2010,Sau2010,Sau2012,Stenger2019,Shach2015} and fractional quantum hall devices~\cite{Camino2007,Goldman1010,Chamon1997}. While trivial-topological phase transition~\cite{xiao2020} and measurement-based braiding~\cite{Wootton2017} have been observed on a (non-topological) quantum computer, thus far there has been no definitive experimental evidence of braiding due to dynamical state evolution~\cite{Bartolomei2020,Camino2007,Ofek2010,McClure2012,Willett2013, Nakamura2019, willett2019}.  

In this work, we simulate the key part of a topological quantum computer: the dynamics of braiding of a pair of MZMs on a tri-junction.
Specifically, we use the results of Ref.~\cite{Backens2017} to map a minimum model of a topological tri-junction~\cite{Alicea2011,Halperin2012,Hyart2013,Heck2012,Hassler2011,Stenger2019} onto a three-qubit Hamiltonian. Braiding is implemented by parametrically adjusting the Hamiltonian parameters; the time evolution is implemented using the Suzuki-Trotter decomposition, with each time step implemented by one- and two-qubit gates. We significantly boost the fidelity of our quantum time-evolution code by using pulse-level control to scale cross resonance (CR) gates~\cite{Chow2011} derived from those pre-calibrated on the backend, thereby enabling coupling of qubit pairs with shorter CR gate times.  Specifically, we observe that using native CNOT gates, we can move a MZM from one arm of the tri-junction device to another arm, thus performing 1/6 of a full braid. However, using the scaled gates, we are able observe a complete braid. We remark that this result can be interpreted as an experimental observation of braiding in a quantum system.

\begin{figure}[t]
\begin{center}
\includegraphics[width=\columnwidth]{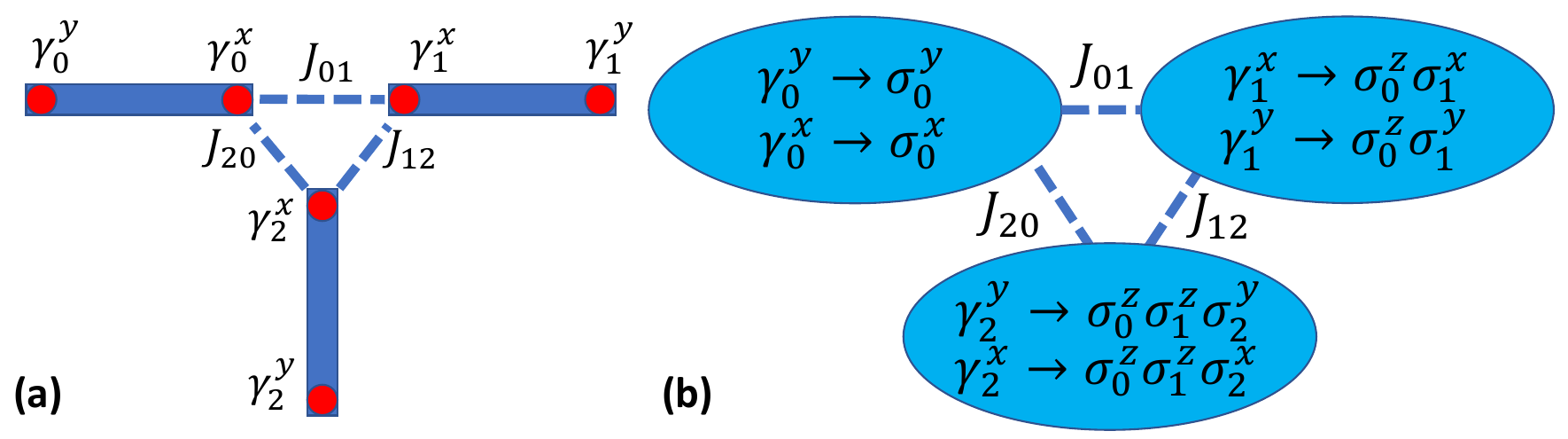}
\end{center}
\vspace{-2mm}

\caption{Topological tri-junctions. (a) Schematic depiction of a tri-junction device as proposed in Refs.~\cite{Alicea2011,Stenger2019,Halperin2012,Hyart2013,Heck2012,Hassler2011}.  The arms of the device are represented by blue bars, Majorana modes by red circles and the tunable couplings between pairs of inner Majorana modes by dashed lines. (b) Minimum model composed of three qubits, each qubit contributes a pair of Pauli operators to the description of the Majoranas.  The representation is written in the blue ovals.}
\label{F1}
\vspace{-3mm}
\end{figure}

\begin{figure}[t]
\includegraphics[width=\columnwidth]{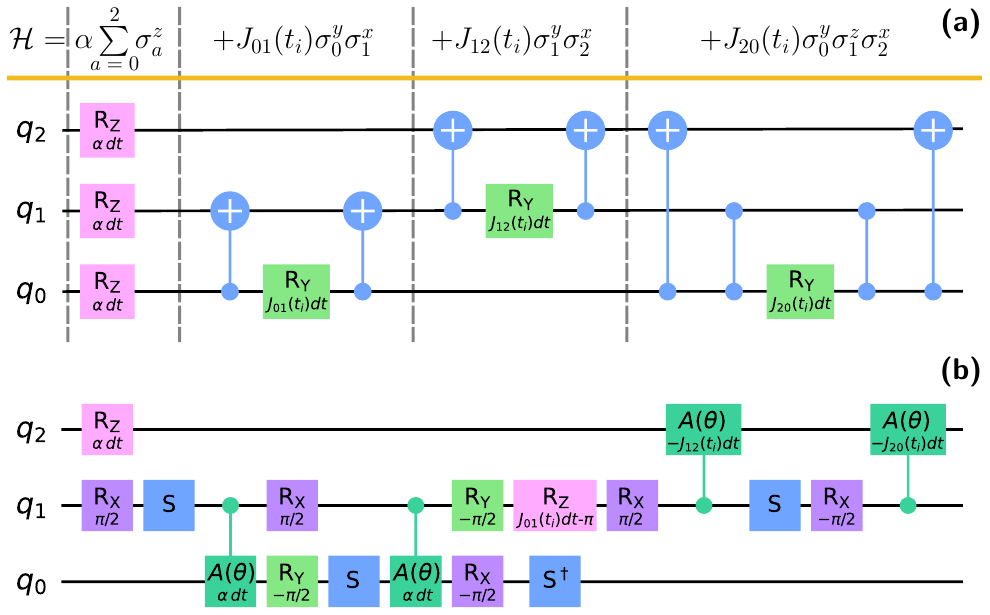}
\caption{Gate sets that implement a single Suzuki-Trotter step $t_i$ of the time evolution. (a) Implementation with 2-qubit basis gates; each set of gates is labeled by the corresponding Hamiltonian terms above. (b) Implementation with analog scaled 2-qubit echoed CR gates $A(\theta)$, where $ZX(\theta) = R_x(\pi/2) A(\theta) R_x(\pi/2)$, as shown in Fig.~\ref{fig:qpt}c.}
\label{F2}
\end{figure}

{\it A minimal model of a topological tri-junction --}
Our model is constructed using the six Majorana operators depicted in Fig.~\ref{F1}a and is described by the Hamiltonian~\footnote{See Section A of the supplement for a model that describes both high energy modes and MZMs and the corresponding quantum code.} 
\begin{equation}
    H=i\alpha\sum_{a=0}^2\gamma_a^x\gamma_a^y+i\frac{1}{2}\sum_{a\neq b}J_{ab}\gamma_a^x\gamma_b^x.
    \label{eq:Hfermion}
\end{equation}
Here, $\gamma_a^x$, $\gamma_a^y$ are Majorana operators on the inner and outer ends of arm $a$ respectively, $\alpha$ describes the coupling between Majorana modes on the same arm, 
and $J_{ab}=-J_{ba}$ couples Majorana modes at the tri-junction.  Initially, we set $J_{01}=J_{\text{max}}$ and $J_{12}=J_{20}=0$. This setting results in $\gamma_0^x$ and $\gamma_1^x$ fusing into one complex fermion and $\gamma_2^x$ and $\gamma_2^y$ fusing into another complex fermion. The low-energy sector is spanned by the operators $\gamma_0^y$ and $\gamma_1^y$, which are the two MZMs that we would like to braid. Braiding is preformed by moving the two MZMs around the arms of the tri-junction, which is accomplished by modulating each $J_{ab}(t)$ as a function of time from $0$ to some maximum value which we call $J_\text{max}$ and then back to $0$ following the protocol enumerated in Table~\ref{T1}.  This is the same braiding procedure that would be used on a real tri-junction~\cite{Stenger2019}. This braiding procedure relies on a separation of timescales: braiding should be slow compared to the timescale of the fused MZMs, but fast compared to the splitting of the two ``free'' MZMs~\footnote{See Section B of the supplement for a discussion of topological protection}.  

\begin{table}
\begin{tabular}{c|c|c}
\hline
step & time & couplings \\
\hline
  1  & $t: 0\rightarrow\tau$ & $J_{01}:J_\text{max} \rightarrow 0$, $J_{12}: 0\rightarrow J_\text{max}$ \\
  2  & $t: \tau\rightarrow 2\tau$ & $J_{12}:J_\text{max} \rightarrow 0$, $J_{20}: 0\rightarrow J_\text{max}$ \\
  3  & $t: 2\tau\rightarrow 3\tau$ & $J_{20}:J_\text{max} \rightarrow 0$, $J_{01}: 0\rightarrow J_\text{max}$ \\
  4-6 & $t: 3\tau\rightarrow 6\tau$ & repeat steps 1-3
\end{tabular}
\caption{Procedure for braiding Majorana zero modes depicted in Fig.~\ref{F1} by modulating the couplings $J_{01}$, $J_{12}$, and $J_{20}$.  Each step moves a Majorana from one arm to another.  In our calculations, we set $J_{\text{max}} = 1$, $\alpha = 3~J_{\text{max}}$, and $\tau = 3.3~1/J_\text{max}$.  After three steps, the Majoranas are swapped, after six steps the Majoranas have returned to their initial positions.}
\label{T1}
\end{table}

To simulate braiding on a quantum computer we map the fermionic Hamiltonian Eq.~\eqref{eq:Hfermion} onto a Hamiltonian 
\begin{equation}
    H(t)=\alpha\sum_{a=0}^2\sigma_a^z+J_{01}(t)\sigma_0^y\sigma_1^x+J_{12}(t)\sigma_1^y\sigma_2^x+J_{20}(t)\sigma_0^y\sigma_1^z\sigma_2^x
    \label{eq:Hboson}
\end{equation}
that acts on qubits. Here, $\sigma_a^{\nu}$ is the Pauli $\nu$-matrix acting on qubit $a$.
The Hilbert space of the minimal model of Eq.~\eqref{eq:Hfermion} is equivalent to the Hilbert space of three qubits, depicted in Fig.~\ref{F1}b. The Majorana operators are related to the qubit operators via the Jordan-Wigner transformation, see Fig.~\ref{F1}(b).

To simulate the dynamics of Eq.~(\ref{eq:Hboson}) we must find the quantum gates that approximate the time evolution operator $U=\mathcal{T}\prod_i\exp{\left(-iH(t_i)dt\right)}$, where $\mathcal{T}$ is the time ordering operator and $dt$ is a small time step. Throughout, we rely on the second order Suzuki-Trotter approximation to decompose the time evolution into manageable pieces. We choose the time step $dt$ to maximize the fidelity of the braiding procedure which is determined by a competition between (1) the Suzuki-Trotter error that is minimized by making a large number of small time-steps $dt$ and (2) the error from the imperfect quantum gates which is minimized by reducing the gate count by making $dt$ large. 

{\it Quantum simulation with basis gates \/ --}
We split the time evolution of a single Suzuki-Trotter step into four parts.  Each part is decomposed into single-qubit rotations labeled $R_{\nu,i}(\theta)$ for a $\theta$-rotation around the $\nu$-axis of qubit $i$, and CNOT gates labeled $C_{x,ij}$, as depicted in Fig.~\ref{F2}a.
We initialize the qubits to
\begin{equation}
    \psi_{\pm}(0)= R_{y,2}(\pi) R_{y,0}(\mp\pi/2) C_{x,01} R_{x,0}(-\pi/2)\ket{000},
\label{psi+}
\end{equation}
where the states $\psi_{\pm}(0) = \ket{\pm} \equiv \ket{e}\pm\ket{g}$ with $\ket{g}$ and $\ket{e}$ being the ground and first excited state of the system corresponding to the even $\ket{e}$ and odd $\ket{g}$ parity states of the two MZMs.  Because there is a finite coupling between the MZMs, given by $\alpha$ in Eq.~\eqref{eq:Hboson}, $\ket{g}$ and $\ket{e}$ are at slightly different energies.  Any state outside of the low energy subspace defined by $\ket{g}$ and $\ket{e}$ can be understood as a linear combination of higher energy states.  After a single braid we expect: $\psi_{\pm}(6\tau)  = \ket{\mp}$. To check if braiding was successful, we invert the initialization gates so that $\ket{000}$ corresponds to successful braiding.

\begin{figure}[t]
\begin{center}
\includegraphics[width=\columnwidth]{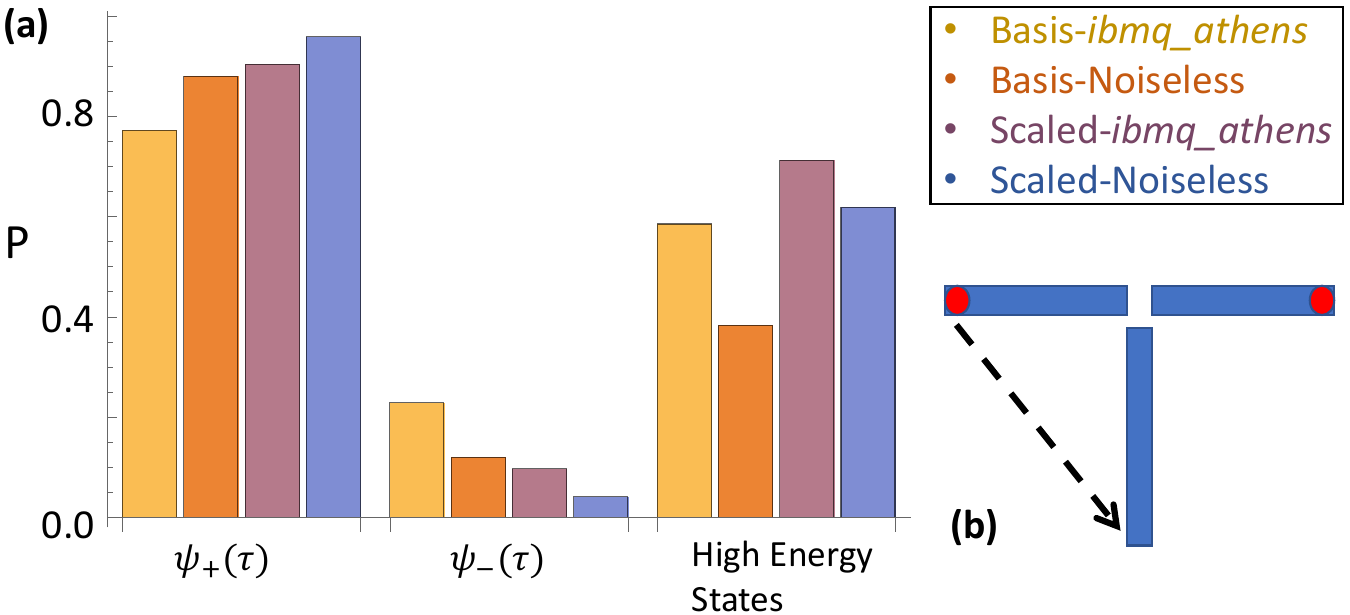}
\end{center}
\vspace{-2mm}
\caption{Panel (a): Comparing the probability distributions obtained using noiseless classical simulations and quantum computer calculations (see legend) obtained after moving a MZM from one arm of a tri-junction to another arm (as depicted in panel (b)). 
The $\psi_{+}(\tau)$ and $\psi_{-}(\tau)$ states form the low-energy subspace, with $\psi_{+}(\tau)$ corresponding to the target state in the absence of time-step and gate errors; in constructing the plot we normalized probabilities in this subspace so that $P_{\psi_{+}(\tau)}+P_{\psi_{-}(\tau)}=1$. The percentage of the counts which end up outside of the computational space is labeled "High Energy States".  A large number of the counts (38\% even in Basis-Noiseless) end up in the high energy states due to the small number of time steps.
}
\label{F3}
\vspace{-3mm}
\end{figure}

A full braid with only three time steps per MZM swap (i.e. $\tau$ is broken into three time steps) requires 96 CNOT gates and cannot be simulated on {\it ibmq\_athens} as the CNOT gates (the dominate source of error) have an error of $1.3\%$ and $0.9\%$ between qubits $(0,1)$ and $(1,2)$, respectively.  The CNOT gate for qubits $(0,2)$ is built out of the other two.  
However, moving a single MZM from one arm of the tri-junction to another arm (thus performing 1/6th of a complete braid),  requires only $3\times4=12$ two qubit gates.  This corresponds to step 1 in table~\ref{T1} with $J_{20} = 0$. The set of gates depicted in Fig.~\ref{F2} was translated into the basis gates using Qiskit's transpile function at optimization level 3 and 1024 shots were measured. Figure~\ref{F3} shows the resulting probability distribution~\footnote{See section C of the supplement for additional data and details of the measurement procedure.}.
The probabilities labeled `noiseless' were calculated using the QASM simulator in Qiskit to perform fully coherent quantum evolution and represent the size of time-step errors.
We observe a clear preference for the $\psi_{+}(\tau)$ state when running on {\it ibmq\_athens}, indicating that we have successfully moved a MZM from one arm of the tri-junction to another. However, when performing the full braid, there is no longer a clear preference as we measure $P_{-}(\tau)/P_{+}(\tau) = 1.04 \pm 0.12$ where $P_{\pm}$ is the probability of system ending in the $\ket{\pm} state$

\begin{figure*}
    \centering
    \includegraphics[width=\textwidth]{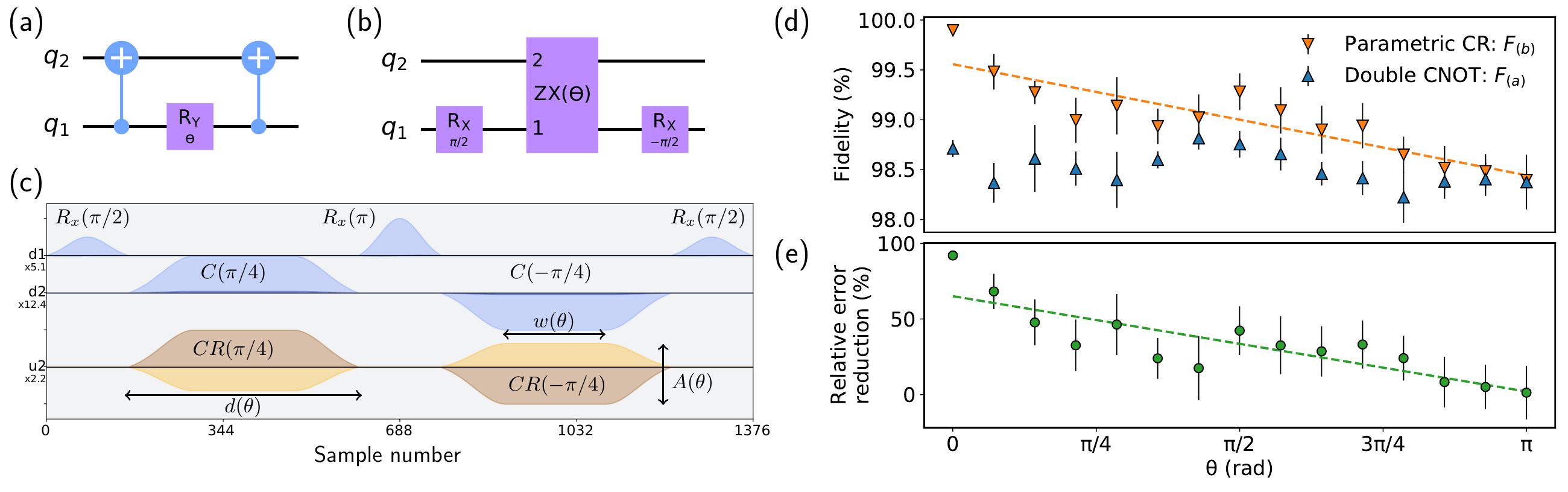}
    \caption{Quantum process tomography of $U_{yx}(\theta)$ as a function of $\theta$ on {\it ibmq\_athens}.  Both the double-CNOT circuit (a) and the parametric $ZX(\theta)$-based circuit (b) create $U_{yx}(\theta)$. 
    The circuit in (b) at $\theta=\pi/2$ is implemented by the pulse schedule shown in (c). Here, each sample lasts $0.222~{\rm ns}$.
    (d) Fidelity $F_{(a)}$ (blue up-triangles) and $F_{(b)}$ (orange down triangles) of quantum circuits (a) and (b), respectively, obtained with QPT.
    (e) Relative error reduction of the pulse-efficient circuit in (b) over the circuit in (a), i.e. $1-E_{(b)}/E_{(a)}$ where $E_{(x)}=1-F_{(x)}$.}
    \label{fig:qpt}
\end{figure*}

{\it Designing scaled quantum gates \/ -- }
We decrease two-qubit gate errors by designing scaled gates which implement smaller rotations in the two-qubit Hilbert subspace than the basis CNOT gates, which apply a full $\pi/2$ rotation in the Hilbert space.  
We focus on the operator $U_{yx}(\theta) = e^{-i \theta\sigma^y_1 \sigma^x_2/2}$, which arises from the Jordan-Wigner transformation, see Fig.~\ref{F2}a.
$U_{yx}(\theta)$ can be implemented with 2 CNOT gates as in Fig.~\ref{fig:qpt}a. The basis CNOT gates are created from $U_{zx}(\pi/2)$ implemented by CR. These CNOTs are echoed CR pulses~\cite{Sheldon2016} on the control qubit with rotary echoes~\cite{Sundaresan2020} on the target qubit, combined with single-qubit gates before and after the echoed CR gate that generate the correct direction of CNOT. Since errors mainly arise during CR pulses, we minimize their duration in the braiding algorithm. This is achieved with pulse-level control enabled by Qiskit Pulse~\cite{McKay2018}.  We implement $U_{yx}(\theta)$ using $U_{zx}(\theta)$ operations (see Fig.~\ref{fig:qpt}b) derived from the highly-calibrated CNOT pulse schedules~\cite{Gokhale2020, Alexander2020}. Modifying the CNOT pulse schedules avoids additional calibrations which is paramount when running jobs through a queue on cloud-based quantum computers \footnote{see section D of the Supplement for details on the implementation of the scaled gates}. The rotation angle $\theta$ depends on the area under the pulses, and is often considerably less than $2 \times \pi/2$ for the Majorana braiding simulation schedule (Fig.~\ref{F2}). This allows us to build $U_{zx}(\theta)$ gates with considerably shorter CR pulses, and hence introduce less error per Suzuki-Trotter step.

To benchmark the $U_{yx}$ analog quantum gate, we compare it to the double-CNOT basis gate implementation (see Fig.~\ref{fig:qpt}a and b). 
We measure the gate fidelity with quantum process tomgraphy (QPT) for 15 values of $\theta$ linearly spaced between 0 and $\pi$.
Each measurement is done with 2048 shots and repeated four times to gather statistics.
We mitigate readout errors by preparing each of the four basis states and measuring the outcome which we use to correct the QPT data~\cite{Bravyi2020b, Barron2020}.
Fidelity measurements of the benchmark circuit are interleaved with those of the scaled CR circuit to mitigate biases in our comparison due to drifts.
We observe that the scaled CR pulses systematically have a higher fidelity than the double-CNOT implementation at all measured values of $\theta$, see Fig.~\ref{fig:qpt}d.
The double-CNOT benchmark should have a constant fidelity as the rotation angle $\theta$ is implemented by a virtual $Z$ gate~\cite{McKay2017} once the schedule is transpiled to {\it ibmq\_athens}.
We therefore attribute the fidelity fluctuations in Fig.~\ref{fig:qpt}d to drifts as the data were acquired over a three-day period. Such fluctuations are also observed in the fidelity of the scaled CR pulses.
We observe that the analog circuit is strongly advantageous to the basis gate circuit at small $\theta$, see Fig.~\ref{fig:qpt}e, which is the relevant case for quantum simulation. We attribute the decrease in fidelity of the analog circuit at larger $\theta$ to an increase in errors caused by the longer duration CR pulses. 

{\it Quantum simulation with scaled gates \/ -- }
The Hamiltonian~\eqref{eq:Hboson} contains a three-qubit coupling term which induces a unitary time evolution that cannot be efficiently encoded using our analog two-qubit quantum gates. We perform one more basis transformation, which results in the three-qubit gate being replaced by a pair of two-qubit gates~\footnote{see section E of the supplement for details of the basis transformation}. The resulting quantum circuit, after hand optimization to combine single-qubit rotations, is depicted in Fig.~\ref{F2}b.

\begin{figure}[t]
\begin{center}
\includegraphics[width=\columnwidth]{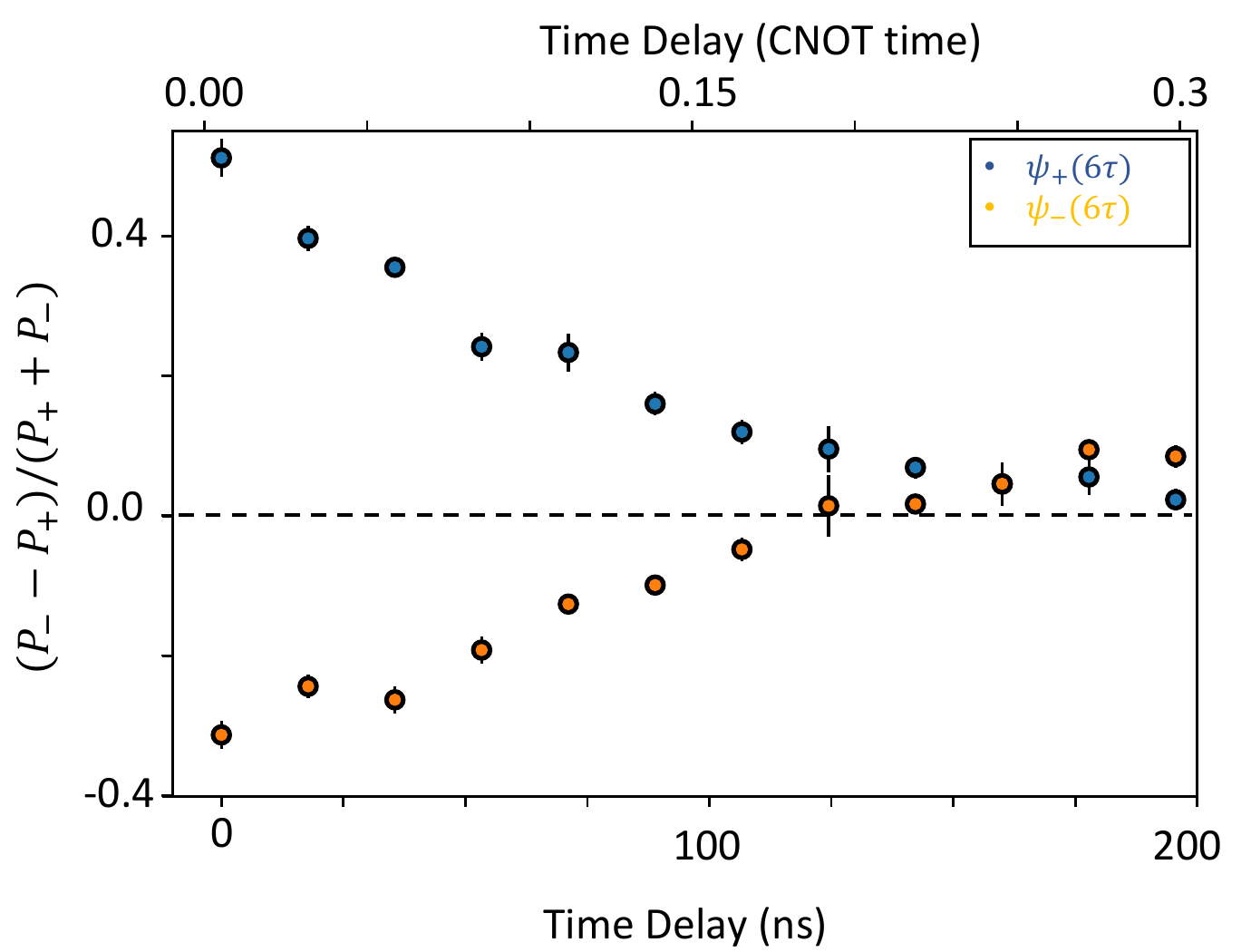}
\end{center}
\vspace{-2mm}

\caption{Bias towards braiding as a function of delay time added to the ZX rotation gate.  We plot $(P_{-}-P_{+})/(P_{-}+P_{+})$ so that 1 (-1) corresponds to perfect braiding for the plus (minus) state and 0 corresponds to an equal distribution of shots in the braiding target state and in the trivial (non-braiding) state.}
\label{F5}
\vspace{-3mm}
\end{figure}

We now return to Fig.~\ref{F3} to compare the fidelity of the  simulations implemented with basis and scaled gates.  We note that quantum simulations performed with basis and scaled gates use a different wave function basis and therefore have slightly different time-step errors as seen from the high energy states in Fig~\ref{F3}.  We observe a larger leakage out of the low energy subspace for the scaled computation than for the basis gates, but the fidelity in the low energy basis is better.  
Crucially, the scaled-gate computation on the quantum computer is much closer to its noiseless ideal than is the basis-gate calculation on the quantum computer to its own noiseless ideal, highlighting the improvement in the fidelity achieved with analog quantum gates.

In Fig.~\ref{F5} we plot the results of running the braiding pulse schedule on {\it ibmq\_athens} with a variable amount of incoherent error added in the form of an additional delay time appended to all 2-qubit gates~\footnote{See Section F of the supplement for additional data}.  The probabilities $P_{+}$ and $P_{-}$ are found by counting the number of shots that end in the $\psi_{+}$ and $\psi_{-}$ states, respectively, after the braiding procedure. Each data point is the average of four trials with 8192 shots in each trial after applying the same readout error mitigation as with the QPT~~\cite{Bravyi2020b, Barron2020}.  We use the ratio $(P_{-}-P_{+})/(P_{+}+P_{-})$ to measure success in braiding, this ratio should be positive if the final state is closer to $\psi_{-}(0)$ and negative if it closer to $\psi_{+}(0)$. In the absence of additional noise, we observe a strong preference for successful braiding. As we introduce additional noise this preference slowly diminishes, disappearing completely when the delay exceeds 150~ns or about $25\%$ of the duration of a CNOT gate. The observation that additional noise washes out the braiding signal (a) supports our interpretation that we are indeed observing a quantum coherent process of braiding and (b) explains why we were unable to observe braiding with CNOT gates which extend the pulse schedule significantly past the 150~ns per two-qubit gate at which the braiding signal disappears~\footnote{See Section G of the supplement for a noise model}.

{\it In summary \/ --} We have demonstrated that pulse-level control of quantum computers enables us to simulate the braiding of Majorana zero modes, thus expanding the ``digital'' capabilities offered by the native basis gates on IBM Quantum backends. These two-qubit operations were derived from the highly-calibrated basis CNOT of the backend and required no further calibration. Our demonstration shows that we have reached the point at which quantum computers can perform interesting quantum simulations, but to achieve a sufficient quality it is crucial to understand the performance of the hardware and compose software that respects its limitations.
Looking towards the future, the ease of programming obtained from digital abstractions is outweighed by the increase in performance obtained by programming the ``analog'' pulse schedules in a noise-aware method.  We therefore argue that the path forward for quantum simulation in the noisy quantum era is the use of abstraction-free programming of quantum computers.  This allows for continuous time evolution on part of the Hamiltonian and could be a path towards fully continuous Hamiltonian evolution.

\section{Acknowledgements}

The authors acknowledge use of the IBM Quantum Experience devices for this work. The authors also thank S.M. Frolov, N. Earnest-Noble, D.T. McClure, N. Kanazawa and E. Pritchett for insightful discussions and G. Ben-Shach for a careful read of the manuscript. J.S. and D.P. acknowledge support from NSF PIRE-1743717.

\newpage

\bibliographystyle{apsrev4-1}
\bibliography{Ref}

\clearpage

\appendix
\section{Quantum computer code for simulating a tri-junction with longer topological superconducting wires}
\label{qccfsatwlt}
\begin{figure}[b]
\begin{center}
\includegraphics[width=\columnwidth]{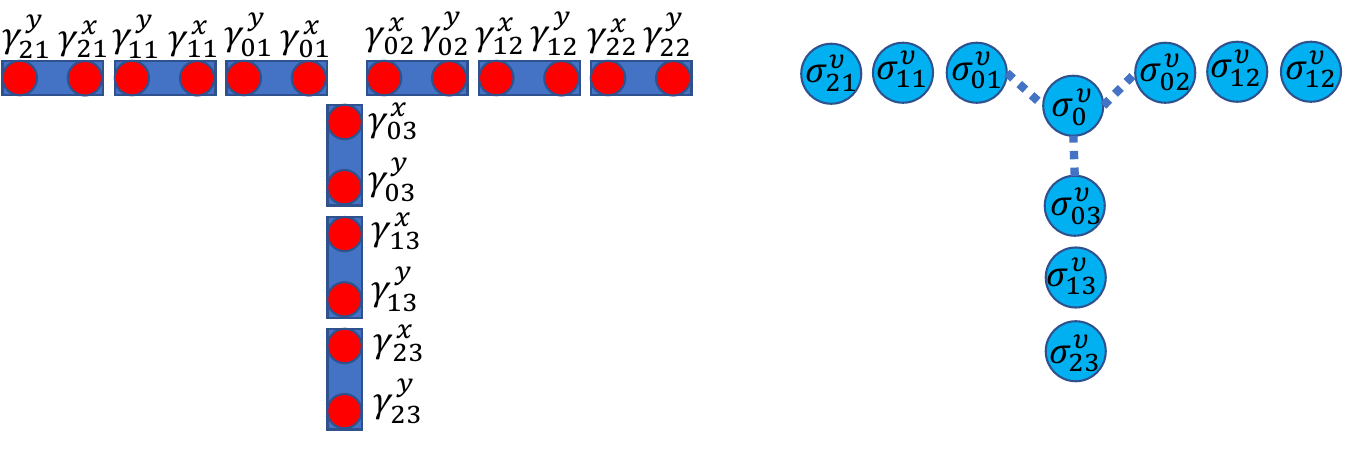}
\end{center}
\vspace{-2mm}

\caption{A ten qubit setup.  On the left we depict the fermion model and on the right the corresponding qubit ``device''.  Note that there is an extra qubit with operators $\sigma_0^{\nu}$ which does not correspond to a site in the fermion model.  This qubit is used to enforce fermion commutation relations while preserving locality~\cite{Backens2017}.}
\label{F6}
\vspace{-3mm}
\end{figure}
The three qubit model is convenient due to its simplicity, however, the overlap of the Majorana modes is controlled by hand through the parameter $\alpha$ in Eq.~\eqref{eq:Hfermion}.  Instead, we would like to control the overlap by the number of qubits which separate the end modes -- thus obtain topological protection.  In this model, every arm of the T-junction is composed of $L$ qubits, see Fig.~\ref{F6}.  The Hamiltonian becomes,
\begin{equation}
    \begin{split}
    H(t)=&i\sum_{a=1}^3\left(\mu\sum_{i=0}^{L-1} \gamma_{ia}^y\gamma_{ia}^x+\Delta\sum_{i=0}^{L-2}\gamma_{ia}^y\gamma_{(i+1)a}^x\right)
    \\
    +&i\frac{1}{2}\sum_{a,b,c}J_{ab}(t)\epsilon_{abc}\gamma_{0a}^x\gamma_{0b}^x, 
    \end{split}
\end{equation}
where $\epsilon_{abc}$ is the Levi-Civita symbol, $\mu$ is the on-site potential, $\Delta$ is the p-wave coupling term between sites, $J_{ab}=J_{ba}$ couples the arms of the tri-junction, and $\gamma_{ia}^{x}$ and $\gamma_{ia}^{y}$ are the two Majorana operators that act on site $i$ of arm $a$.  In other words, we have three Kitaev chains coupled at the first site of each chain.  If $\mu$, $\Delta$, and $J_{ab}(t)$ are held constant, then increasing $L$ acts to increase the length of the wire and so one can study how topological protection increases as length increases.  

We can model each fermion operator as a string of spin operators acting on qubits:
\begin{equation}
    \begin{split}
        &\gamma_{ia}^x=\sigma_0^a\prod_{j=0}^{i-1}\sigma_{ja}^z\sigma_{ia}^x,
        \\
        &\gamma_{ia}^y=\sigma_0^a\prod_{i=0}^{i-1}\sigma_{ja}^z\sigma_{ia}^y,       
    \end{split}
\end{equation}
where $\sigma_{ia}^{\nu}$ is the Pauli $\nu$-matrix acting on the qubit labeled $ia$. The auxiliary qubit with Pauli matrix $\sigma_0^a$ is added to keep string operators local \cite{Backens2017}. In terms of these spin operators, the Hamiltonian is
\begin{equation}
    \begin{split}
    H(t)=&\sum_{a=1}^3\left(\mu\sum_{i=0}^{L-1}\sigma_{i,a}^z+\Delta\sum_{i=0}^{L-2}\sigma_{i,a}^x\sigma_{i+1,a}^x\right)
    \\
    +&\frac{1}{2}\sum_{a \neq b \neq c}J_{ab}(t)\sigma_{0,a}^x\sigma_{0,b}^x \sigma_0^c.
    \end{split}
\label{A3}
\end{equation}
Notice that the auxiliary qubit $\sigma^c_0$ appears only in the junction coupling term.

Once again we write the sums in the exponential for the time evolution operator $U=\mathcal{T}\prod_i\exp{\left(-iH(t_i)dt\right)}$ as products of exponentials.  In this way, time evolution is defined by the gate sets
\begin{equation}
    e^{-i\mu\sigma_{ia}^z dt/2} = R_{z,ia}(\mu dt)
\end{equation}
\begin{equation}
    e^{-i\Delta\sigma_{ia}^x\sigma_{(i+1)a}^x dt/2} = C_{x,ia,(i+1)a} R_{x,ia}(\Delta dt) C_{x,ia,(i+1)a}
\end{equation}
\begin{equation}
    \begin{split}
    &e^{-iJ_{ab}(t)\sigma_{0,a}^x\sigma_{0,b}^x \sigma_0^c dt/2}
    \\
    &=C_{c,0,0a} C_{x,0b,0a} R_{x,0a}(J_{ab}(t) dt) C_{x,0b,0a}C_{c,0,0a}        
    \end{split}
\end{equation}
where $C_{\nu,\alpha,\beta}$ is a controlled $\nu\in\{x,y,z\}$ gate acting on the qubits labeled by $\alpha$ and $\beta$.

The qubits are initialized to the two degenerate ground states of the Hamiltonian written in the Jordan-Wigner basis (Eq.~\eqref{A3}) for $\mu=0$, $J_{12}=J_{20}=0$, and $J_{01}=\Delta$. These ground states can be reached from the empty state $\psi_0=\ket{000\ldots}$ by applying a particular gate set.  First we initialize the auxiliary third arm of the tri-junction:
\begin{equation}
\begin{split}
    &\psi_{a=3}=
    \\
    &C_{x,23,23}R_{y,23}\left(\frac{\pi}{2}\right)R_{x,13}(\pi)C_{x,03,13}R_{y,03}\left(\frac{\pi}{2}\right)R_{x,0}(\pi)\psi_{0}.
\end{split}
\end{equation}
This is the same for both ground states.  They can be reached by:
\begin{equation}
\begin{split}
        \psi_{\pm}(0)=&\ldots R_{y,12}\left(\mp\frac{\pi}{2}\right)R_{y,02}\left(\pm\frac{\pi}{2}\right)
        \\
        &\ldots R_{y,11}\left(\mp\frac{\pi}{2}\right)R_{y,01}\left(\pm\frac{\pi}{2}\right)\psi_{a=3}
\end{split}
\label{psi10}
\end{equation}
If we start in the $\psi_{+}(0)$ state and apply the conjugate transpose of the gate set that defines $\psi_{-}(0)$ at the end of the braiding procedure then $\psi_0$ corresponds to successful braiding. 

Although there are currently quantum devices with enough qubits to simulate the nine fermion device, the extra two qubit gates (6 per time step) make this simulation impractical. However, as gate fidelities improve checking the topological protection of Majoranas on the nine fermion device will become possible.

\section{Hints of topological protection from the three qubit model}

\begin{figure}[t]
\begin{center}
\includegraphics[width=\columnwidth]{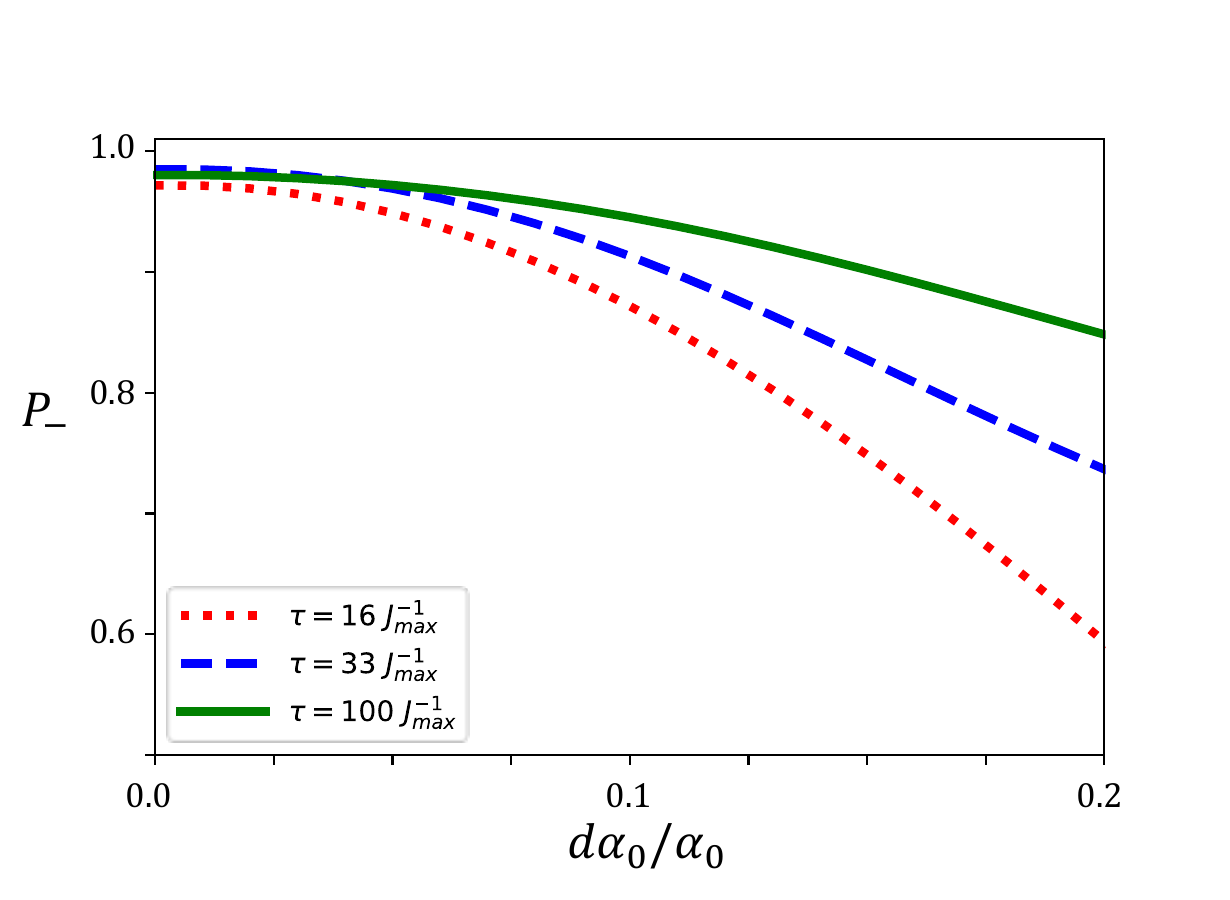}
\end{center}
\vspace{-2mm}

\caption{Braiding fidelity as a function of a local perturbation. The probability $P_-$ to be in the target quantum state is shown against a shift $d\alpha_0$ of parameter $\alpha_0 = \alpha+d\alpha_0$ on qubit 0 where $\alpha_1=\alpha_2=\alpha$.  The fidelity is plotted as a function of the change of $\alpha_0$ on the first qubit to the optimized $\alpha_0$.  The protection of the braiding procedure depends on the protocol time $\tau$.  The parameter $\alpha$ is optimized for each protocol time.}
\label{FE}
\vspace{-3mm}
\end{figure}

Topological protection from braiding manifests itself as a tolerance to local perturbations. However, braiding MZMs at a finite rate results in the topological protection being imperfect. In our case the braiding time is dictated by the separation of time scales associated with the strength of the MZM couplings in the tri-junction. To study tolerance to local perturbations we rewrite the Hamiltonian in Eq.~(1) of the main text so that each arm of the tri-junction has a unique coupling~$\alpha_a$.
\begin{equation}
    H=i\sum_{a=0}^2\alpha_a\gamma_a^x\gamma_a^y+i\frac{1}{2}\sum_{a\neq b}J_{ab}\gamma_a^x\gamma_b^x
\end{equation}
In Fig.~\ref{FE} we plot the braiding fidelity as a function of a small perturbation to $\alpha_0$.  As the procedure becomes more adiabatic, the system becomes more protected.  Each curve in the figure corresponds to a different protocol time $\tau$.  As $\tau$ increases the curve flattens out indicating a higher tolerance to local perturbations.  
This is consistent with the expectation that there is additional protection to local perturbations as the MZM couplings in the tri-junction arms become weaker, e.g. as the number of qubits that compose each arm of the tri-junction increase.  

To truly get the benefit of topological protection, one must use a tri-junction model with arms composed of multiple qubits each coupled by local parameters.  The parameter $\alpha_0$ can be thought of as a coarse graining of these local parameters.   The system becomes truly topologically protected only as $\alpha_0$ approaches zero.   In this case, each step of the braiding procedure is achieved by tuning the local parameters so that the edge between topological and trivial superconductors (qubits) can slide towards the tri-junction, the tri-junction couplings are adjusted, and then the topological-trivial edge is brought back out to the the end of the tri-junction arm. 

\section{Tracking the wave function trajectory during braiding}

\begin{figure}[b]
\begin{center}
\includegraphics[width=\columnwidth]{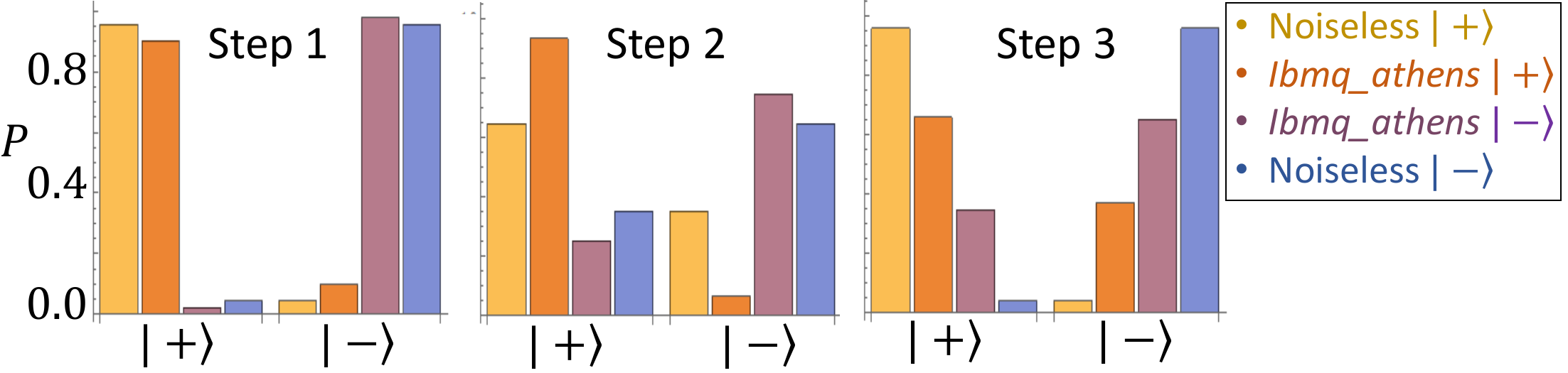}
\end{center}
\vspace{-2mm}

\caption{Probability distributions in the computational subspace during the first three steps of the braiding procedure.  The yellow (Noiseless) and orange ({\it ibmq\_athens}) bars are for the state which originates in the $\ket{+}$ state and the purple (Noiseless) and blue ({\it ibmq\_athens}) bars are for the state which originates in the $\ket{-}$ state.  The labels on the horizontal axis indicate where each state would end up during each step of the process if there was no trotterization error.  }
\label{Track}
\vspace{3mm}
\end{figure}

The first three steps of the braiding procedure swap the two MZMs.  While the fidelity of the swap is characterized by the target state on the third step, it is instructive to do an experiment which ends at each of the three steps to see that the states are following the correct path.  To do this we unwind the basis in which braiding occurs so that the target state after each step is mapped onto a specific occupation state of the device.  This is accomplished by applying
\begin{equation}
    \begin{split}
        R_{y,2}\left(\frac{\pi}{2}\right) R_{x,1}\left(\frac{\pi}{2}\right) R_{y,0}\left(\frac{\pi}{2}\right) & \text{ after step 1,}
        \\
        R_{y,2}\left(\frac{\pi}{2}\right) R_{y,1}\left(\frac{\pi}{2}\right) R_{x,0}\left(\frac{\pi}{2}\right) & \text{ after step 2, and}
        \\
        R_{y,0}\left(\frac{\pi}{2}\right) & \text{ after step 3.} 
    \end{split}
\end{equation}
Steps 4 and 5 require several CNOT gates to unwind the basis, therefore, we end this analysis at step 3.  The $\ket{+}$ and $\ket{-}$ states stay separated throughout the process, as seen by the separation of probability distributions in Fig.~\ref{Track}. This observation supports our interpretation that the quantum device is truly following the expected Hamiltonian evolution and is not ending in the correct state by other means.

\section{Implementation with Qiskit Pulse}
IBM quantum computers provide highly calibrated single-qubit and CNOT gates~\cite{Jurcevic2020}.
Virtual single-qubit $Z$-gates are created by phase shifting subsequent microwave drives~\cite{McKay2017}.
In a basis gate quantum computation where the only two-qubit operation is the CNOT gate, the operator $U_{yx}(\theta)=e^{-i\theta\sigma^y_1\sigma^x_2/2}$ is built with an $R_{y,1}(\theta)$ rotation sandwiched between two CNOT gates, see Fig.~\ref{fig:qpt}a of the main text. However, the same operation can be implemented in a pulse-efficient manner by moving the parameter $\theta$ into a single two-qubit CR gate which ideally implements $U_{zx}(\theta)=e^{-i\theta\sigma^z_1\sigma^x_2/2}$  \cite{Chow2011}.
The required circuit is thus $R_{x,1}(\pi/2)\cdot U_{zx}(\theta)\cdot R_{x,1}(-\pi/2)$, see Fig.~\ref{fig:qpt}b  of the main text.

Pulse-level control of IBM backends is enabled by Qiskit Pulse \cite{McKay2018, Alexander2020}.  The CNOT basis gates are built from  echoed CR pulses which consist of the $CR(\pm\pi/4)$ pulses sandwiching an $X$-rotation echo pulse applied to the control qubit to cancel undesired $ZI$ and $IX$ terms of the CR Hamiltonian~\cite{Sheldon2016}.  The compensation rotary pulses $C(\pm\pi/4)$ applied to the target qubit suppress the remaining non-commuting $ZZ$ and $IY$ terms \cite{Sundaresan2020}.

The $CR(\pm \pi/4)$ pulses are calibrated to have the shortest duration while minimizing leakage outside of the computational basis, as determined by randomized benchmarking~\cite{Sheldon2016}. The effect of decoherence is thus minimized and the pulse amplitude that retains the qubits in the computational subspace is maximized. Since the largest errors arise when performing CR pulses we wish to minimize their duration in the braiding algorithm. This is achieved by (a) using the $U_{zx}(\theta)$ gate to implement $U_{yx}(\theta)$ instead of two CNOT gates and (b) by creating the pulse schedules for $U_{zx}(\theta)$ by modifying the highly-calibrated CNOT pulse schedules \cite{Gokhale2020, Alexander2020}. The rotation angle $\theta$ depends on the area under the pulses, and is often considerably less than $2 \times \pi/2$ for the Trotterization that braids MZM. This allows us to build $U_{zx}(\theta)$ gates with considerably shorter duration CR pulses than two CNOT gates, hence introducing less error per Trotter step.  Since $CR(\pm\theta)$ and $C_\pm$ are all implemented with flat-top Gaussian pulses and that $\theta$ depends non-linearly on the pulse amplitude $A$ \cite{Magesan2020} we modify the pulse area by first stretching and compressing the width $w$ of the flat-tops.
We only scale the amplitude when $w=0$.
This avoids any additional calibration as the relation $\theta(w)$ is linear.
Avoiding additional calibration on cloud-based quantum computers is paramount when running jobs through a queue.
The $CR(\pm\pi/4)$ pulses are given in the parametric form \texttt{GaussianSquare}$(d, A, \sigma, w)$ with an area given by 
\begin{align}
    \alpha^*=|A^*|w^*+|A^*|\sigma\sqrt{2\pi}\,\text{erf}(n_\sigma).
\end{align}
Here, $n_\sigma$ is the number of standard deviations $\sigma$ contained in the pulse with total duration $d$ and flat-top width $w$ and amplitude $A$.
The quantities $d$, $w$, and $\sigma$ are all specified in units of Arbitrary Waveform Generator (AWG) samples which last $0.222~{\rm ns}$ on {\it ibmq\_athens}.
The star superscript indicates that we are referring to the parameters of the CNOT schedule.
To scale the CR gates we first calculate the target area of each pulse
\begin{align}
    \alpha(\theta)=\frac{\theta}{\pi/2}\alpha^*.
\end{align}
As long as $\alpha(\theta)>|A^*|\sigma\sqrt{2\pi}\,\text{erf}(n_\sigma)$ we change the width of the pulse following
\begin{align}\label{eqn:width}
    w(\theta)=\frac{\alpha(\theta)}{|A^*|}-\sigma\sqrt{2\pi}\,\text{erf}(n_\sigma).
\end{align}
When $\alpha(\theta)<|A^*|\sigma\sqrt{2\pi}\,\text{erf}(n_\sigma)$, i.e. when the flat-top vanishes, we instead scale the amplitude of the remaining Gaussian pulse according to
\begin{align}\label{eqn:amp}
    |A(\theta)|=\frac{\alpha(\theta)}{\sigma\sqrt{2\pi}\,\text{erf}(n_\sigma)}.
\end{align}
The phase of the pulse, i.e. $\arg(A)$, is left unchanged to implement the $\pm$ rotations in the echo.
Since the AWGs can only load pulses if their duration is a multiple of $m=16$ samples we set the duration of our pulses to
\begin{align}\label{eqn:duration}
    d = \left \lceil \frac{w(\theta)+n_\sigma\sigma}{m}\right \rceil m\quad\text{samples}.
\end{align}
The pulse schedule implementing $U_{yx}(\theta)$, shown in Fig.~\ref{fig:qpt}c  of the main text, has three single-qubit pulses.
The first pulse is the $R_{x,1}(\pi/2)$ seen in Fig.~\ref{fig:qpt}b  of the main text.
The second pulse is the $R_{x,1}(\pi)$ needed in the echoed CR gate.
The third pulse is $R_{x,1}(\pi/2)$ which corresponds to the second $R_{x,1}(\pi)$ pulse in the echoed CR gate together with the $R_{x,1}(-\pi/2)$ seen in Fig.~\ref{fig:qpt}b  of the main text. 

\begin{figure*}[t]
\begin{center}
\includegraphics[width=\textwidth]{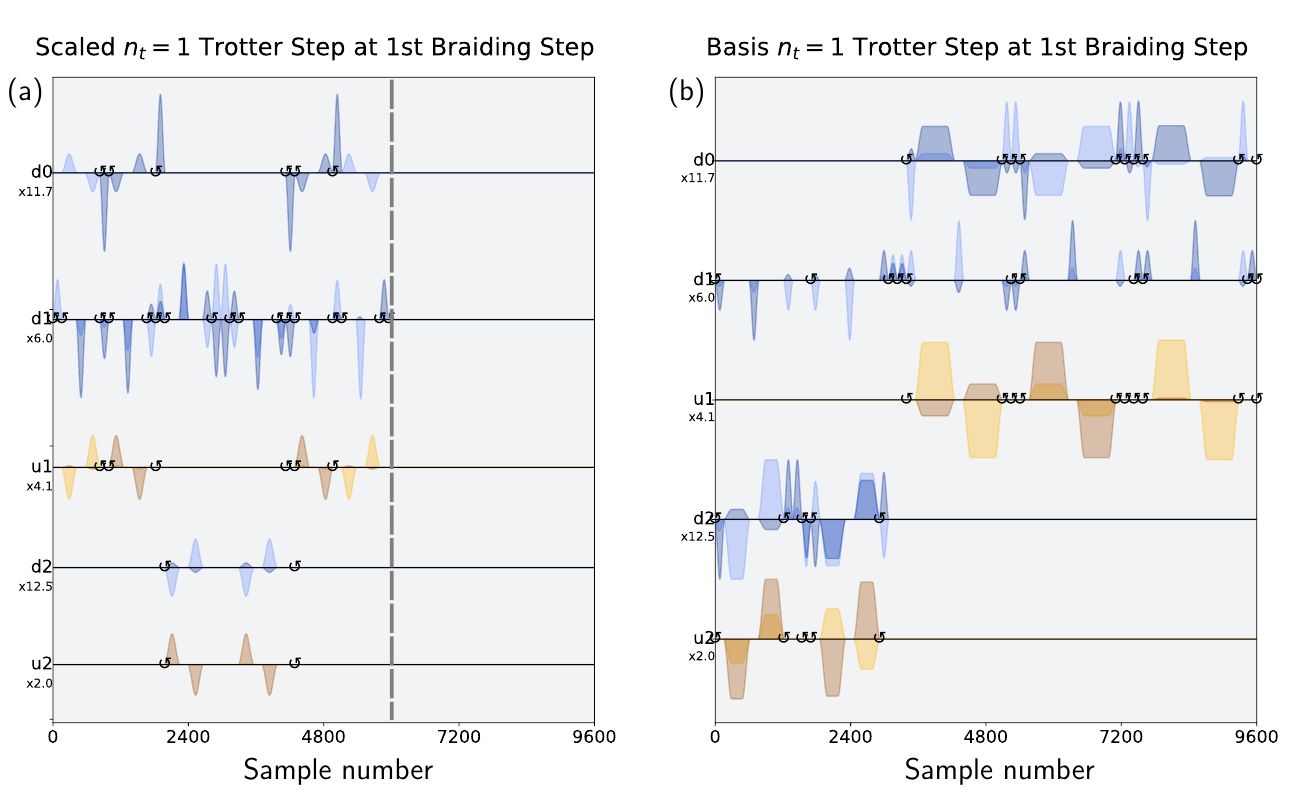}
\caption{Pulse schedules corresponding to the (a) scaled circuit and (b) basis circuit, corresponding to gates given in Fig.~\ref{F2}, with the end of the scaled schedule marked by the vertical gray dashed line. Here, the `d' and `u' correspond to the (on-resonant) \texttt{DriveChannel}s and (off-resonant) \texttt{ControlChannel}s of Qiskit Pulse, respectively. The pulses are shaded according to phase, with light pulses in-phase and dark pulses in the quadrature phase. The circular arrows correspond to $Z$-rotations executed in software by changing the phase of subsequent pulses~\cite{McKay2017}. The duration is given in units of sample time, $dt = 0.222$~ns. 
\label{fig:schedules}}
\end{center}
\end{figure*}

A side-by-side comparison of the pulse schedules generated by a highly-transpiled (optimization level 3) circuit using the basis CNOTs to generate $U_{yx}(\theta)$ interactions as in Fig.~\ref{fig:qpt}a and the scaled pulses of Fig.~\ref{fig:qpt}b-c is presented in Fig.~\ref{fig:schedules} to highlight their differences. These pulse schedules correspond to the circuits of Fig.~\ref{F2} for a single Trotter step, in this case the second ($n_t=1$) step of the first Braiding step, as $J_{01} \rightarrow 0$ and $J_{12} \to J_{\rm max}$. The duration of the scaled circuit is $5952/9200 = 62\%$ of the basis circuit, leading to a reduction in the amount of decoherence that would occur regardless of the applied pulses. Furthermore, the bulk of errors on current noisy quantum hardware occur during the two-qubit operations, as depicted by the yellow pulses on the \texttt{ControlChannel}s in both schedules. The area of the two-qubit pulses is substantially less when scaling the CR pulses than when using CNOT gates which indicates why the scaled pulses yield a successful MZM braiding that is impossible to observe with CNOT-based circuits. For the Trotter step in Fig.~\ref{fig:schedules} the ratio of areas shows that the scaled CR is $463.51/1544.87 = 30\%$ that of the basis CNOTs.

\section{Rotated basis}
The Hamiltonian of the three dot system (Eq.~(2) in the main text) has a thee-qubit coupling term $J_{20}(t)\sigma_0^y\sigma_1^z\sigma_2^x$
whose evolution operator cannot be continuously generated from scaled CR gates.  We therefore rotate the Hamiltonian into a basis where there are no three-qubit coupling terms.  

Take $U=(a+b)/\sqrt{2}$ as a general unitary operator.  We want $U$ to transform all terms in the Hamiltonian into terms with less than three qubits.  That is $U^{\dagger}\sigma_0^y\sigma_1^z\sigma_2^x U$, $U^{\dagger}\sigma_1^y\sigma_2^x U$,  $U^{\dagger}\sigma_0^y\sigma_1^x U$, and $ U^{\dagger}\sigma_a^z U$ must contain fewer than three Pauli matrices for all $a$.  Take $U=(a+b)/\sqrt{2}$ as a general unitary operator and $\lambda$ to be a general Hermitian operator. Then we have that
\begin{equation}
     U^{\dagger}\lambda U =  \begin{cases} 
      \lambda & \text{if} \quad a \lambda (\lambda a)^{-1} = +b \lambda (\lambda b)^{-1} \\
      a b \lambda & \text{if} \quad a \lambda (\lambda a)^{-1} = -b \lambda (\lambda b)^{-1}
   \end{cases} 
\end{equation}
In other words, we need to find an $a$ and $b$ such that $a$ and $b$ have different commutation relations with $\sigma_0^y\sigma_1^z\sigma_2^x$ and $ab\lambda$ is a one or two qubit operator.  Additionally, for all other operators in the Hamiltonian $\lambda^{\prime}$, we need that either $ab\lambda^{\prime}$ is a one or two qubit operator or for $a$ and $b$ to have the same commutation relation with $\lambda^{\prime}$.  

Notice that $a=\sigma^z_1$ and $b=\sigma^y_0 \sigma^x_1$ have all of the above properties.  Applied to the qubit Hamiltonian (Eq.~(2) of the main text), we get:
\begin{equation}
\begin{split}
    \Bar{H}(t)&=(\sigma^z_1+\sigma^y_0 \sigma^x_1)H(t)(\sigma^z_1+\sigma^y_0 \sigma^x_1)/2 
    \\
    &=\alpha\left(\sigma_0^x\sigma_1^y+\sigma_0^y\sigma_1^x+\sigma_2^z\right)
    \\
    &+J_{01}(t)\sigma_1^z+J_{12}(t)\sigma_1^y\sigma_2^x+J_{20}(t)\sigma_1^x\sigma_2^x.
\end{split}
\end{equation}
We have exchanged the three-qubit term for two two-qubit terms.  Since there are more multi-qubit terms overall, there is no reason to expect that this basis will have less error if the braiding procedure is implemented using basis gates.  However, since there are no three-qubit gates, we can simulate the entire braiding process using the scaled CR gates.  

To initialize the state in this basis we use the initialization in Eq.~\eqref{psi+} of the main text and apply the basis rotation gate set:
\begin{equation}
    R_{z,1}(\pi) C_{y,01} R_{y,0}(-\pi/2)C_{y,01}
\end{equation}

The last step to implementing the Hamiltonian evolution in this basis is to translate the evolution operators into quantum gates.  To do this, we Trotterize the evolution operator and use the generalized unitary-to-gate transcription.
\begin{equation}
    \begin{split}
        &e^{-i\phi\sigma^a_i/2}=R_{a,i}(\phi)
        \\
        &e^{-i\phi\sigma^a_i\sigma^b_j/2} = C_{a,ij}R_{b,j}(\phi)C_{a,ij}
    \end{split}
\end{equation}
which holds for $b \neq z$ which is always the cases in this basis. Alternatively, we can use the scalled $ZX(\theta)$ gate described in the main text.
\begin{equation}
\begin{split}
    &e^{-i\phi\sigma^a_i\sigma^b_j/2} =
    \\
    &R_{bj}(\pi/2)R_{ai}(\pi/2)ZX(\theta)R_{ai}(-\pi/2)R_{bj}(-\pi/2)
\end{split}
\end{equation}

\section{Noise induced drift in the braiding experiment}

\begin{figure}[t]
\begin{center}
\includegraphics[width=\columnwidth]{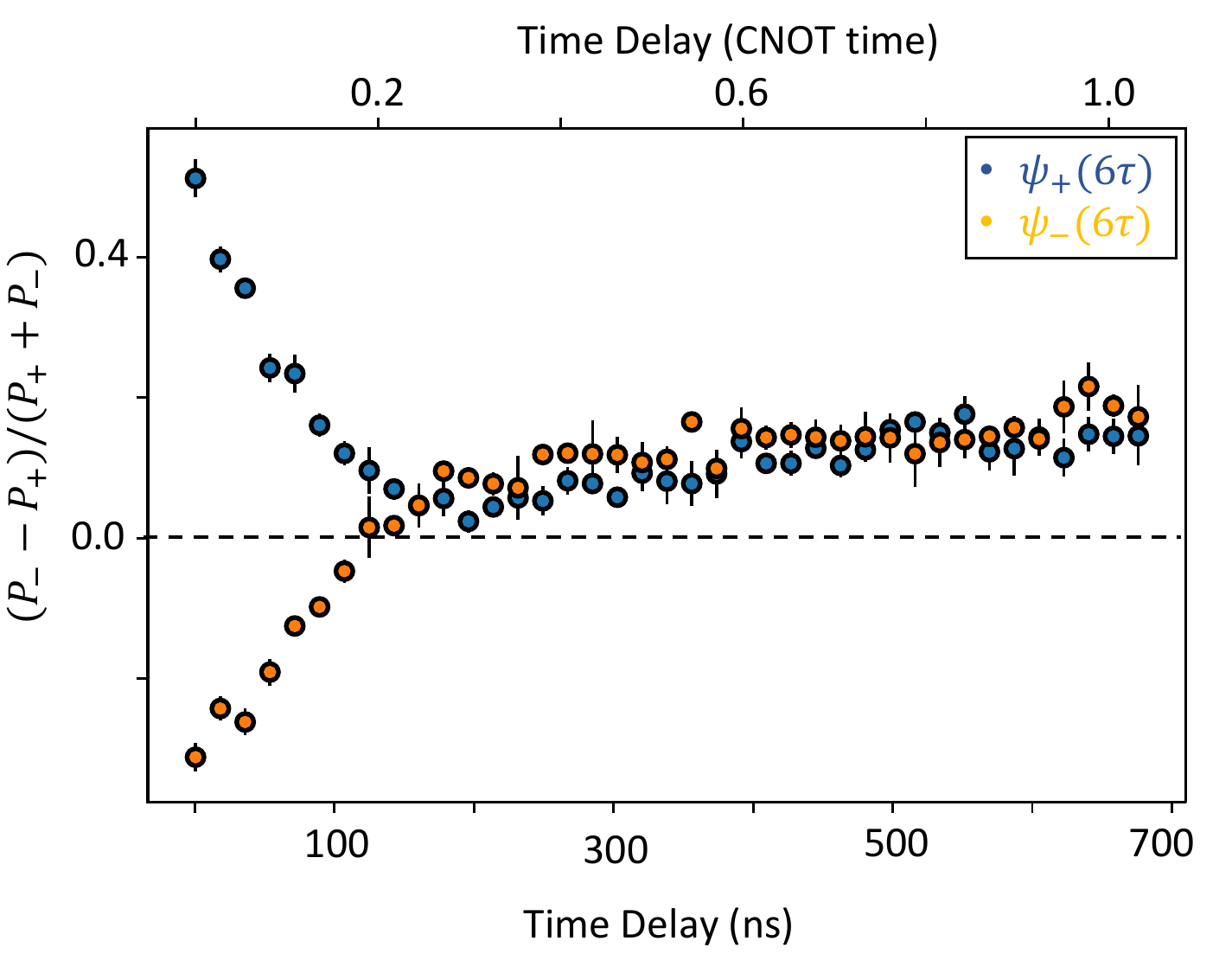}
\end{center}
\vspace{-2mm}

\caption{Bias towards braiding as a function of delay time added after each $ZX$ rotation.  Similar to Fig.~5 in the main text but the data has been extended to longer time delays.}
\label{SupTrack}
\vspace{3mm}
\end{figure}

In Fig.~\ref{SupTrack} we plot all of the data accumulated to construct Fig.~5 of the main text, including data at large delay times which was cut off in the figure in the main text as it shows no signs of braiding.  The data is acquired in sequential order with all of the low time-delay data taken before the long time-delay data. 
When the delay is greater than $\sim 200~{\rm ns}$ the probability to end up in $\ket{+}$ or $\ket{-}$ is essentially independent of the initial state, indicating that the added noise overwhelms the quantum braiding dynamics. We attribute the delay dependency of $(P_- - P_+)/(P_+ + P_-)$ to drifts in the quantum device over the course of the experiment.  Similar drifts are also seen in Fig.~4d of the main text, where the fidelity of the double-CNOT circuit oscillates with $\theta$.

\section{Error Model}

\begin{figure}[b]
\begin{center}
\includegraphics[width=\columnwidth]{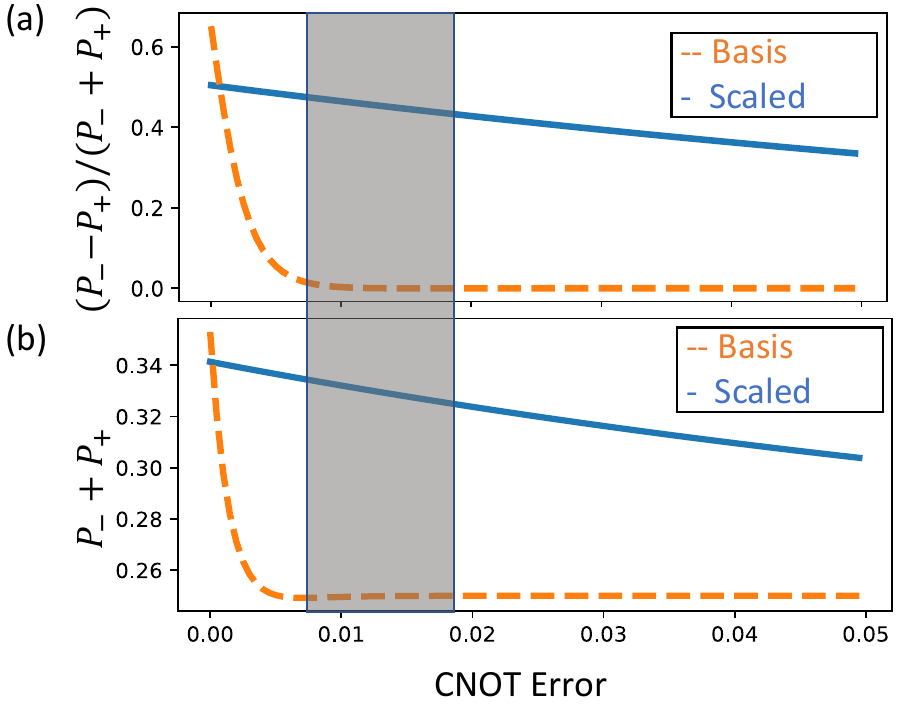}
\end{center}
\vspace{-2mm}

\caption{(a) Ratio of the probability of successfully braiding to the total probability of being in the computational space.  (b) Probability of being in the computational space. The blue and orange curves show the scaled CR gates and the basis CNOT gates, respectively.  The gray shaded region is the typical range of CNOT errors in real devices.   While the basis gate procedure has little or no bias towards successfully braiding, the scaled CR procedure is clearly biased towards successfully braiding.  The difference in fidelity ratio at zero CNOT error is due to an increased number of single-qubit gates in the scaled CR procedure.  For these plots, we have $\alpha = 0.2~J_{\rm max}$ and $\tau = 3.3~J_{\rm max}$.  We find that the best balance between Suzuki-Trotter error and gate error is to perform each protocol in thee time steps.}
\label{F4}
\vspace{-3mm}
\end{figure}

To illustrate the advantage of scaled gates over basis ones for quantum simulation we construct a simple error model that describes errors accumulated in the CR procedure.  The error model consists of single bit flip errors which are proportional to the duration of the CR gate, described by the superoperator 
\begin{equation}
\begin{split}
    &\text{Error}_1 = (1-\epsilon) \text{SuperOp}(\hat{I})+\epsilon \text{SuperOp}(\hat{X})
    \\
    &\text{Error}_2 = \text{Error}_1 \otimes \text{Error}_1
\end{split}
\end{equation}
where $\text{Error}_1$ refers to a single-qubit error and $\text{Error}_2$ refers to a two-qubit error, $\text{SuperOp}(\hat{O})$ denotes the superoperator of operator $\hat{O}$ and $\epsilon$ is determined from the device.  

For the {\it ibmq\_athens} device, the single-qubit gate errors range from $2.2\times 10^{-4}$ to $2.8\times 10^{-4}$. For the CR error, we take the CNOT error and scale it to the phase shift we want to apply.  In other words, $\epsilon = (\phi/\pi) \epsilon_{CNOT}$, were $\epsilon_{CNOT}$ is the CNOT error and $\phi$ is the desired rotation angle.  The CNOT errors in the {\it ibmq\_athens} device range from $6.9\times 10^{-3}$ to $9.4\times 10^{-3}$.  In Fig.~\ref{F4}, we plot probabilities for both the basis gate procedure using controlled gates and the scaled procedure using CR gates for a range of CNOT errors.  The top panel plots the probability of braiding if we project onto the computational basis $(P_+-P_-)/(P_++P_-)$ while the bottom panel shows the probability of being in the the computational basis $P_++P_-$.  The blue curve is generated using the scaled CR procedure while the orange curve is for the basis gate procedure.  Typical values of the CNOT error are shaded in gray.  There is no bias for the basis gate procedure in the shaded region, but there is a bias for the scaled CR gates.   For the real pulse experiment, the braiding probability (Fig.~5 of the main text) is in the expected range, however, there is only a small bias for the computational subspace ($P_+ + P_- = 0.28\pm 0.01$) which suggest that there are sources of error that we do not include in our model.

\end{document}